# Graphyne- and Graphdiyne-based Nanoribbons: Density Functional Theory Calculations of Electronic Structures

L. D. Pan, L. Z. Zhang, B. Q. Song, S. X. Du, [a] and H.-J. Gao [a]

*Institute of Physics, Chinese Academy of Sciences, Beijing 100190, P R China*

Abstract

We report on the configurations and electronic properties of graphyne and graphdiyne nanoribbons with armchair and zigzag edges investigated with first principles calculations. Our results show that all the nanoribbons are semiconductors with suitable band gaps similar to silicon. And their band gaps decrease as widths of nanoribbons increase. We also find that the band gap is at the Γ point for all graphdiyne ribbons and it is at the X point for all graphyne ribbons. Of particular interest, the band gap of zigzag graphyne nanoribbons show a unique "step effect" as the width increases. This property is good for tuning of the energy band gap, as in a certain range of the ribbon width, the energy gap remains constant and in reality the edge cannot be as neat as that in a theoretic model.

---

[a] Author to whom correspondence should be addressed.   E-mail: sxdu@iphy.ac.cn

Graphene, a single atomic layer of carbon sheet, has attracted tremendous attention because of its unique properties.[1-5] Graphdiyne, a new carbon allotrope, which has the same symmetry as graphene and has butadiyne linkages between its nearest-neighbor hexagonal rings,[6,7] has recently synthesized and fabricated on copper, showing experimentally the semiconductor property with conductivity of $2.516 \times 10^{-4}$ S m$^{-1}$, which is comparable to silicon.[8] This work inspires researchers in this field to explore new possible applications in future electronics. It is well known that band gap engineering of graphene includes the graphene nanoribbon widths and edge functionalization, etc.[9,10] Similarly, these methods of tuning band gap are expected to be applicable to graphdiyne. In addition, graphdiyne nanoribbons have much larger natural "holes" than graphene, thus it is easier to realize doping of various candidates into the "holes" to fabricate *n*-doping or *p*-doping one dimensional semiconductor. Graphyne is of another structure with large "holes", which belongs to the same family as graphdiyne. It is obtained by replacing one third of the carbon-carbon bonds in graphene with acetylenic linkages.[11] Work on synthesis of both graphyne and graphdiyne substructures has been successfully conducted due to promising electrical and optical properties for future electronics.[12] It is well known that the energy gap of quasi-one-dimensional (1D) graphene nanoribbons (GNR) is tunable and shows unique electronic properties.[13,14] However, little information about the electronic properties of 1D graphyne or graphdiyne nanoribbons has been reported up to date.

In this Letter, we focus on tuning electronic structure and band gap by the nanoribbon widths. We give out configurations and electronic properties of graphyne/graphdiyne nanoribbons, both with zigzag edges and armchair edges. The results show that all these nanoribbons are semiconductors with controllable band gaps of 0.5 ~ 1.3 eV, depending on the width and the edge morphology. It is also found out that, unlike GNRs, the band gaps of zigzag graphyne nanoribbons show a unique "step effect."

All calculations are performed within density functional theory (DFT) and the local density approximation (LDA),[15] using the Vienna *ab-initio* simulation package (VASP).[16,17] The projector augmented wave (PAW) method[18,19] is employed. The electronic wave functions are expanded in plane waves with a kinetic energy cutoff of 400 eV. A vacuum layer of 15 Å is applied in the direction perpendicular to the ribbon plane and between the neighboring ribbons. The k-points mesh used in the calculations is 1 x 5 x 1, generated automatically with the origin at the Γ point. Other choices of k points are tested, and the energy difference is found to be less than 0.005 eV. The structures are relaxed until residual forces are smaller than 0.01 eV/Å.

In order to obtain the nanoribbon structures, we optimize the two-dimensional (2D) structures of graphyne and graphdiyne, getting lattice constants of 6.83 Å and 9.38 Å, respectively. Cutting through an infinite graphyne sheet along two directions indicated as X and Y in Fig. 1(a), we obtain armchair and zigzag graphyne nanoribbons of various widths. There are two kinds of terminations – benzene ring and acetylene. To simplify the calculation and considering the stability of nanoribbon structures, we investigate only the nanoribbons with benzene ring terminations. Graphdiyne nanoribbons are obtained in a similar way. The dangling σ bonds at the edges are saturated with hydrogen in the present work. Then all these nanoribbons with various edges and widths are further optimized. Figures 1(b) and 1(c) show the configurations of armchair and zigzag graphyne nanoribbons, respectively, where $n$ indicates the number of repeating units. Unlike armchair nanoribbons, in zigzag nanoribbon structures, the number of repeating units $n$ can differ by a half-integer. In our calculation, the width ranges from $n = 1.5$ to $n = 4$ for zigzag graphyne nanoribbons and from $n = 2$ to $n = 8$ for armchair graphyne nanoribbons. Graphdiyne nanoribbons can easily be obtained from graphyne nanoribbons, merely by replacing one acetylenic linkage with two acetylenic linkages between two neighboring benzene rings.

It was reported that three-dimensional graphyne is either semiconducting or metallic depending on its stacking arrangement.[7] Two-dimensional sheets of graphyne and graphdiyne are both semiconductors with band gaps of 0.52~0.53 eV.[6] In our work, 26 kinds of nanoribbons are investigated and the results demonstrate that all of them are semiconductors with different band gaps varying by widths and edge types. Band gaps of graphyne based nanoribbons are 0.59~1.25 eV for armchair ones and 0.75~1.32 eV for zigzag ones. For graphdiyne nanoribbons, the gaps are 0.54~0.97 eV for armchair nanoribbons and 0.73~1.65 eV for zigzag ones. Previous simulation of graphene nanoribbons employing LDA method showed that graphene nanoribbons with armchair or zigzag edges both have non-zero direct band gaps varying with their width,[10] whereas the gaps of graphene nanoribbons are generally smaller than those of graphyne ribbons with a similar width. For example, the band gap of an armchair graphyne nanoribbon with a width of 20 Å is about 0.8 eV, whereas that of an armchair graphene nanoribbon with the same width is about 0.5 eV. The band gaps of graphyne and graphdiyne nanoribbons, ~1 eV, is similar to silicon. This excellent semiconductor feature is useful for future application in the field of electronics and semiconductors.

Besides the gap tuning by widths, we find an interesting rule: band gap emerges at the boundary of the BZ (X point) for graphyne nanoribbons while it emerges at the Γ point for graphdiyne nanoribbons. The band gaps of two-dimensional (2D) structures follow the same rule: the band gap lies at the M point for a graphyne sheet and at the Γ point for graphdiyne in reciprocal space. It is also reported that for an even number of acetylenic linkages the band gap is at the Γ point and for an odd number the gap is at the M point, which is at the boundary of the Brillouin Zone (BZ). This rule is verified in cases of up to 4 acetylenic linkages.[6] To confirm the rule, we investigate various nanoribbons, which can be classified into four groups by their edge types (zigzag or armchair) and the number of acetylenic linkages (graphyne or graphdiyne). We chose one configuration from each group and their band structures and density of states (DOS) are shown in Figs. 2(a)-(d) for armchair graphyne nanoribbons ($n=3$), armchair graphdiyne nanoribbons ($n=3$), zigzag graphyne nanoribbons ($n=3$), zigzag graphdiyne nanoribbons ($n=3$), respectively. These results show that the band gaps of graphyne nanoribbons are at the X point and those of graphdiyne nanoribbons are at the Γ point. It is found that this characteristic shows no width and edge dependence for all graphyne and graphdiyne nanoribbons we studied. Therefore, this rule should be an inherent characteristic of graphyne-based and graphdiyne-based carbon allotropes.

As the gap emerges at Γ point in BZ for graphdiyne based structure, its gap tuning is dominated by states neighboring Γ point and less influenced by BZ boundary states. Comparatively, gap tuning of graphyne based structure is mainly influenced by BZ boundary states. Tailing graphyne/graphdiyne into ribbons causes hexagonal BZ reducing to one-dimensional BZ, namely, a "loss" of states near BZ boundary. Increasing ribbon widths can be considered as a process of recovering the states near boundary of BZ. Therefore, the gap of the graphyne nanoribbon can have a subtle dependence on ribbon width. From Figs. 2(e) and 2(f), we can see that gaps of these nanoribbons are clearly width dependent: the band gap of the nanoribbon decreases as the width increases. The decreasing tendency is not surprising, because the ribbon will return to a two-dimensional structure if the width extends infinitely, which is expected to have a band gap of 0.5 eV.[6] In Fig. 2(e), we find that the red one, which indicates the band gaps of zigzag graphyne nanoribbons, reveals an interesting "step effect." In contrast, the black curve, standing for the band gaps of armchair graphyne nanoribbons, has a relatively smooth decline. Figure 2(f) shows the width-dependence of the band gap of a graphdiyne nanoribbon. Note that neither zigzag nor armchair graphdiyne nanoribbons show a "step effect". We can use graphyne-$n$ to indicate the graphdiyne family sturctures, where $n$ indicates the number of triple bonds between the neighboring carbon hexagons. Remembering the gap location of graphdiyne family depends on the parity of $n$, we expect that for the nanoribbon with even $n$, the gap emerges at Γ point of BZ, whereas for graphyne-$n$ nanoribbon with odd $n$, the gap emerges at boundary of BZ. In addition, step effect can happen in odd $n$ graphyne-$n$ nanoribbons.

In order to reveal the bonding type within these nanoribbons, we project the density of states (PDOS) onto the atomic orbitals ($s$, $p_x$, $p_y$ and $p_z$) of the carbon atoms with different bonding situations. Figures 3 shows the zigzag graphyne nanoribbons ($n$=2) for example. Each of the carbon atoms in the nanoribbons exists in one of three different bonding situations – the edge (C1), binding with an acetylenic linkage (C2) or residing within an acetylenic linkage (C3) (Fig. 3(a)). The PDOSs (Fig. 3(b)-3(d)) show that the electronic states near Fermi energy are mainly contributed by the $p_z$ orbital, so the bonding is mainly π type. This phenomenon is general for other graphyne and graphdiyne nanoribbons.

We also investigate the electron density distribution of the valence band and the conduction band for each configuration. As examples, Figure 4 shows the electron density distribution of armchair graphyne nanoribbon ($n$=3, Fig. 4(a) and 4(b)) and zigzag graphyne nanoribbon ($n$=2, Fig. 4(c) and 4(d)). With regard to the valence band (Figs. 4(a) and 4(c)), the electrons are localized around the benzene ring and the triple bond separately. Therefore, in this system the benzene ring is just like isolated benzene. With regard to the conduction band (Figs. 4(b) and (d)), the π orbitals are not fully delocalized around the aromatic ring. In fact, it overlaps with orbitals of the neighboring atom on the acetylenic linkage more than that of its peer atoms on the aromatic ring.

In summary, the graphyne and graphdiyne nanoribbons exhibit semiconductor properties, with band gaps of 0.59~1.25eV (armchair), 0.75~1.32eV (zigzag) for graphyne nanoribbons, and of 0.54~0.97eV (armchair), 0.73~1.65eV (zigzag) for graphdiyne nanoribbons. The band gaps of these nanoribbons decrease as the widths of the nanoribbons increase. It was further found out that the zigzag graphyne nanoribbons show a "step effect" instead of a smooth gap decrease with increasing width. The band gap occurs at boundary of BZ or $\Gamma$ point is determined by the number of acetylenic linkages between nearest-neighboring hexagonal rings. The large and adjustable band gaps show a great potential of applications in the future electronic devices.

**Acknowledgements.**

This work was partially supported by the Natural Science Foundation of China (NSFC), the MOST "973" projects of China, the Chinese Academy of Sciences (CAS) and Shanghai Supercomputer Center.

**Figure captions**

FIG. 1. Armchair- and zigzag-edged nanoribbons are obtained by cutting through an infinite graphene sheet (a) along two directions X and Y. (b) Armchair-edged graphyne nanoribbons with widths $n$=1, 2 and 8, where $n$ denotes the number of chains of hexagonal carbon rings. (c) Zigzag edged graphyne nanoribbons with widths $n$=1, 1.5, 2 and 4.

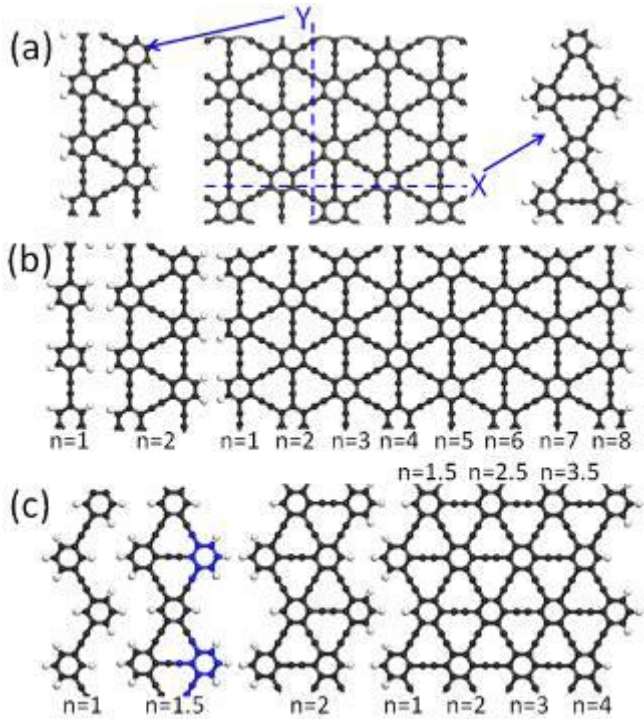

FIG. 2. The band structure and density of states of (a) armchair-edged graphyne nanoribbon ($n$=3); (b) armchair-edged graphdiyne nanoribbon ($n$=3); (c) zigzag-edged graphyne nanoribbon ($n$=3); and (d) zigzag-edged graphdiyne nanoribbon ($n$=3). (e and f) Gaps of graphyne nanoribbons and of graphdiyne nanoribbons with various widths, respectively.

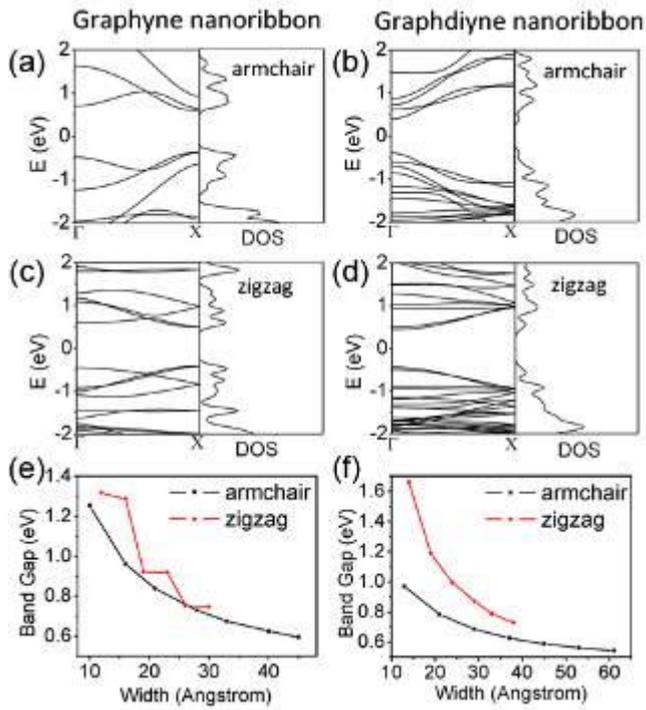

FIG. 3. Density of states of zigzag-edged graphyne nanoribbon ($n=2$) projected on carbon atoms at different locations: (a) One unit cell of zigzag-edged graphyne nanoribbon ($n=2$), (b) C1 (c) C2 and (d) C3.

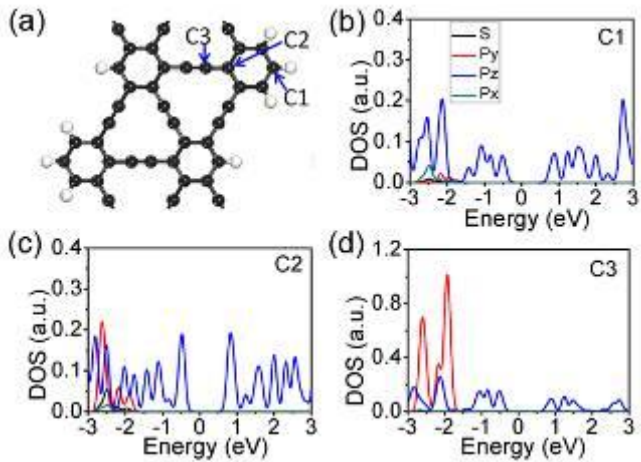

FIG. 4. The charge density distribution near the Fermi level of (a) the valence band of armchair-edged graphyne nanoribbon ($n=3$); (b) the conduction band of armchair edged graphyne nanoribbon ($n=3$); (c) the valence band of zigzag edged graphyne nanoribbon ($n=2$); (d) the conduction band of zigzag-edged graphyne nanoribbon ($n=2$). (Isovalue=0.0069 e/A$^3$)

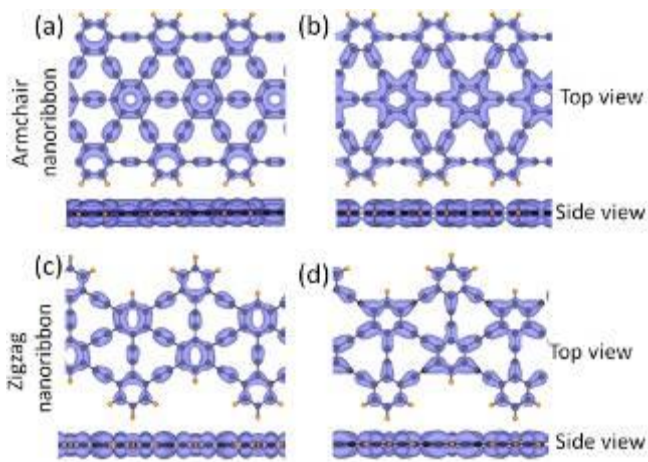